\documentclass[12pt]{article}
\usepackage{amsmath,amssymb}
\newcommand{\eqdef}{\stackrel{\text{def}}{=}}
\newcommand{\n}{\nonumber \\}
\newcommand{\bm}{\boldsymbol}
\newcommand{\ignore}[1]{}
\newcommand{\cF}{c_{\text{\tiny$\mathcal{F}$}}}

\allowdisplaybreaks[4]

\begin{document}

\baselineskip=20pt

\newfont{\elevenmib}{cmmib10 scaled\magstep1}
\newcommand{\preprint}{
   \begin{flushleft}
     \elevenmib Yukawa\, Institute\, Kyoto\\
   \end{flushleft}\vspace{-1.3cm}
   \begin{flushright}\normalsize \sf
     DPSU-10-3\\
     YITP-10-64\\
     {\tt arXiv:1007.3800[math-ph]}\\
   \end{flushright}}
\newcommand{\Title}[1]{{\baselineskip=26pt
   \begin{center} \Large \bf #1 \\ \ \\ \end{center}}}
\newcommand{\Author}{\begin{center}
   \large \bf S.~Odake${}^a$ and R.~Sasaki${}^b$ \end{center}}
\newcommand{\Address}{\begin{center}
     $^a$ Department of Physics, Shinshu University,\\
     Matsumoto 390-8621, Japan\\
     ${}^b$ Yukawa Institute for Theoretical Physics,\\
     Kyoto University, Kyoto 606-8502, Japan
   \end{center}}
\newcommand{\Accepted}[1]{\begin{center}
   {\large \sf #1}\\ \vspace{1mm}{\small \sf Accepted for Publication}
   \end{center}}

\preprint
\thispagestyle{empty}
\bigskip\bigskip\bigskip

\Title{A new family of shape invariantly deformed Darboux-P\"oschl-Teller
potentials with continuous $\ell$}
\Author

\Address

\begin{abstract}
We present a new family of shape invariant potentials which could be
called a ``continuous $\ell$ version" of the potentials corresponding
to the exceptional ($X_{\ell}$) J1 Jacobi polynomials constructed recently
by the present authors. In a certain limit, it reduces to a continuous
$\ell$ family of shape invariant potentials related to the exceptional
($X_{\ell}$) L1 Laguerre polynomials. The latter was known as one example
of the `conditionally exactly solvable potentials' on a half line.
\end{abstract}

\section{Introduction}
\setcounter{equation}{0}

We will present a new family of shape invariant \cite{genden}, thus exactly
solvable, potentials in one dimensional quantum mechanics. The inventory
of exactly solvable quantum mechanics in one dimension \cite{infhul,susyqm}
has seen a rapid increase recently, thanks to the discovery of infinitely
many shape invariant potentials connected with the {\em exceptional\/}
Laguerre, Jacobi, continuous Hahn, Wilson and Askey-Wilson polynomials by
the present authors \cite{os16,os18,os19,hos,stz,os17,os20}.
The exceptional orthogonal polynomials are a new type of orthogonal
polynomials satisfying second order differential (difference) equations.
The $\ell$-th ($\ell=1,2,\ldots$) member of these families of orthogonal
polynomials are sometimes called  $X_{\ell}$ polynomials. They start with
the degree $\ell$ instead of a degree zero constant term, which is the
case for the ordinary orthogonal polynomials. Therefore they are not
constrained by Bochner's theorem \cite{bochner}.
The concept of the exceptional orthogonal polynomials was introduced by 
G\'omez-Ullate et al \cite{gomez}, and the explicit examples of the $X_1$
Laguerre and Jacobi polynomials were constructed within the framework of
the Sturm-Liouville theory. Then Quesne \cite{quesne} reformulated them 
in the language of quantum mechanics. These are the first members
of the infinite families of the exceptional Laguerre and Jacobi
polynomials \cite{os16}. Later another set of $X_2$ Laguerre polynomials
was found \cite{quesne2}, which was generalised to another family of
$X_{\ell}$ Laguerre and Jacobi polynomials \cite{os19}.

Roughly speaking, we are going to derive a ``continuous $\ell$ version" of
the potentials corresponding to the $X_{\ell}$ Jacobi polynomials. The
prepotentials of the exceptional ($X_{\ell}$) Jacobi polynomials are
obtained by deforming those for the Darboux-P\"oschl-Teller (DPT) \cite{dpt}
potential in terms of a degree $\ell$ Jacobi polynomial of twisted
parameters \cite{os16,os19}.
As is well known, the Jacobi polynomials can be expressed in terms of
a Gauss hypergeometric function, which is well defined for non-integer
$\ell$, too. The new family of potentials are obtained by deforming the
DPT potential in terms of the hypergeometric function, which would reduce to 
the Jacobi polynomial for integer $\ell$.
There are, in fact, two types of exceptional Jacobi polynomials, called
J1 and J2 \cite{os19,hos,stz}. It turns out that only the first type,
the J1, deformations give rise to non-singular and shape invariant
potentials. Naturally, the corresponding eigenfunctions are {\em no longer
polynomials\/}. They are a ``continuous $\ell$ version" of the J1 type
$X_{\ell}$ polynomials, $\{P_{\ell,n}\}$. This is in good contrast to the
most known cases of shape invariant potentials, in which the eigenfunctions
are polynomials.

It is well known that the Laguerre polynomials (confluent hypergeometric
functions) are obtained from the Jacobi polynomials (hypergeometric
functions) in a certain limit \cite{szego}. Likewise the exceptional L1
and L2 Laguerre polynomials are derived from the exceptional J1 and J2
Jacobi polynomials, respectively, in the same limit \cite{os19}. This would 
mean that a ``continuous $\ell$ version" of the potentials corresponding to
the L1 $X_{\ell}$ Laguerre polynomials can be obtained from the above
``continuous $\ell$ version" of the potentials corresponding to the
J1 $X_{\ell}$ Jacobi polynomials. In fact, this ``continuous $\ell$ version"
of the potentials corresponding to the L1 $X_{\ell}$ Laguerre polynomials
was derived by Junker and Roy \cite{junkroy} as one example of
`conditionally exactly solvable potentials' in the context of supersymmetric
quantum mechanics. Somehow erroneously this type of potentials had been
declared non-shape invariant \cite{junkroy,duttaroy}. This is partly
because the structure of the corresponding Hamiltonians (potentials)
were not fully understood.
The structure of the Hamiltonians (potentials) of the exceptional orthogonal
polynomials are essentially the same for the J1 $X_{\ell}$ Jacobi and the
L1 $X_{\ell}$ Laguerre polynomials, as shown in our previous papers
\cite{os16,os19,os18,hos,stz}. Therefore we will present the ``continuous
$\ell$ versions" of the potentials corresponding to the J1 $X_{\ell}$ Jacobi
and the L1 $X_{\ell}$ Laguerre polynomials in parallel.
We will follow the notation of \cite{hos,stz}.

This paper is organised as follows. In section two we will recapitulate the
original systems, that is, the quantum mechanical systems of the DPT and
the radial oscillator potentials, in order to set the stage and to introduce
appropriate notation. In section three the new deforming functions, the
``continuous $\ell$ versions" of the deforming polynomials $\xi_{\ell}(\eta)$,
are introduced and their properties are demonstrated. Section four is the
main part of this paper. In subsection 4.1, the deformed systems, that is,
the ``continuous $\ell$ versions" of the potentials corresponding to the
J1 $X_{\ell}$ Jacobi and the L1 $X_{\ell}$ Laguerre polynomials are presented.
Here we stress two points. Firstly, we show the concrete structure of the
Hamiltonians. Secondly, the shape invariance is demonstrated explicitly.
We will briefly mention why the ``continuous $\ell$ versions" of the
potentials corresponding to the J2 $X_{\ell}$ Jacobi and the L2 $X_{\ell}$
Laguerre polynomials do not exist.
The Darboux-Crum transformations \cite{darboux,crum} intertwining the
Hamiltonians of the original systems with the new deformed systems are
introduced in subsection 4.2. Various properties of the new deformed
systems are derived from those of the original systems through the
intertwining relations. In subsection 4.3 we briefly review the limiting
procedure, from the Jacobi polynomials (hypergeometric function) to the
Laguerre polynomials (confluent hypergeometric function), which connects
the results of the ``continuous $\ell$ versions" of the potentials
corresponding to the J1 $X_{\ell}$ Jacobi and the L1 $X_{\ell}$ Laguerre
polynomials. The final section is for a summary and comments.
It is shown why the above recipe to construct a ``continuous $\ell$ version"
does not work for the Hamiltonians of the exceptional Askey type polynomials 
constructed by the present authors \cite{os17,os20}.

\section{Original systems}
\label{sec:org.sys}
\setcounter{equation}{0}

Here we summarise various properties of the original Hamiltonian systems,
the two well known shape invariant systems, the
Darboux-P\"{o}schl-Teller \cite{dpt} and the radial oscillator
\cite{infhul,susyqm} potentials. The eigenfunctions are described by the
Jacobi and Laguerre polynomials, to be abbreviated as J and L.
These results are to be compared with the specially modified systems to
be presented in \S\,\ref{sec:deformed_system}.
Let us start with the Hamiltonians, Schr\"{o}dinger equations and
eigenfunctions ($x_1<x<x_2$):
\begin{align}
  &\mathcal{H}(\bm{\lambda})\eqdef
  \mathcal{A}(\bm{\lambda})^{\dagger}\mathcal{A}(\bm{\lambda}),\quad
  \mathcal{A}(\bm{\lambda})\eqdef
  \frac{d}{dx}-\partial_xw_0(x;\bm{\lambda}),\quad
  \mathcal{A}(\bm{\lambda})^{\dagger}
  =-\frac{d}{dx}-\partial_xw_0(x;\bm{\lambda}),
  \label{origham}\\
  &\mathcal{H}(\bm{\lambda})\phi_n(x;\bm{\lambda})
  =\mathcal{E}_n(\bm{\lambda})\phi_n(x;\bm{\lambda})\quad
  (n=0,1,2,\ldots),\\
  &\phi_n(x;\bm{\lambda})=\phi_0(x;\bm{\lambda})P_n(\eta(x);\bm{\lambda}),
  \quad \phi_0(x;\bm{\lambda})=e^{w_0(x;\bm{\lambda})}.
\end{align}
Here $\eta(x)$ is the sinusoidal coordinate, $\bm{\lambda}$ is the set of
parameters, $w_0(x;\bm{\lambda})$ is the prepotential and
$\mathcal{E}_n(\bm{\lambda})$ is the $n$-th energy eigenvalue:
\begin{align}
  &\eta(x)\eqdef\left\{
  \begin{array}{llll}
  \cos 2x,&x_1=0,&x_2=\tfrac{\pi}{2},&:\text{J}\\
  x^2,&x_1=0,&x_2=\infty,&:\text{L}
  \end{array}\right.\!\!,
  \quad
  \bm{\lambda}\eqdef\left\{
  \begin{array}{lll}
  (g,h),&g,h>0&:\text{J}\\
  g,&g>0&:\text{L}
  \end{array}\right.\!\!,
  \label{eta,lambda}\\
  &w_0(x;\bm{\lambda})\eqdef\left\{
  \begin{array}{ll}
  g\log\sin x+h\log\cos x&:\text{J}\\[1pt]
  -\frac12x^2+g\log x&:\text{L}
  \end{array}\right.\!\!,
  \quad
  \mathcal{E}_n(\bm{\lambda})\eqdef\left\{
  \begin{array}{ll}
  4n(n+g+h)&:\text{J}\\
  4n&:\text{L}
  \end{array}\right.\!\!.
  \label{w0}
\end{align}
The eigenfunction consists of an orthogonal polynomial
$P_n(\eta;\bm{\lambda})$, a polynomial of degree $n$ in $\eta$,
($P_n(\eta;\bm{\lambda})=0$ for $n<0$):
\begin{equation}
  P_n(\eta;\bm{\lambda})\eqdef\left\{
  \begin{array}{ll}
  P_n^{(g-\frac12,h-\frac12)}(\eta)&:\text{J}\\[1pt]
  L_n^{(g-\frac12)}(\eta)&:\text{L}
  \end{array}\right.\!\!.
\end{equation}
Shape invariance \cite{genden} means in this setting \cite{os4,os13,os14}
\begin{equation}
  \mathcal{A}(\bm{\lambda})\mathcal{A}(\bm{\lambda})^{\dagger}
  =\mathcal{A}(\bm{\lambda+\bm{\delta}})^{\dagger}
  \mathcal{A}(\bm{\lambda}+\bm{\delta})
  +\mathcal{E}_1(\bm{\lambda}),\quad
 \bm{\delta}\eqdef\left\{
  \begin{array}{ll}
  (1,1)&:\text{J}\\
  1&:\text{L}
  \end{array}\right.\!\!,
  \label{shapeinv}
\end{equation}
or equivalently,
\begin{equation}
  \bigl(\partial_xw_0(x;\bm{\lambda})\bigr)^2
  -\partial_x^2w_0(x;\bm{\lambda})
  =\bigl(\partial_xw_0(x;\bm{\lambda}+\bm{\delta})\bigr)^2
  +\partial_x^2w_0(x;\bm{\lambda}+\bm{\delta})
  +\mathcal{E}_1(\bm{\lambda}).
  \label{shapeinv2}
\end{equation}
It is straightforward to verify this for the given forms of the
prepotential $w_0(x;\bm{\lambda})$ \eqref{w0}.
The action of $\mathcal{A}(\bm{\lambda})$ and
$\mathcal{A}(\bm{\lambda})^{\dagger}$ on the eigenfunction is
\begin{align}
  &\mathcal{A}(\bm{\lambda})\phi_n(x;\bm{\lambda})
  =f_n(\bm{\lambda})
  \phi_{n-1}\bigl(x;\bm{\lambda}+\bm{\delta}\bigr),
  \label{Aphi=fphi}\\
  &\mathcal{A}(\bm{\lambda})^{\dagger}
  \phi_{n-1}\bigl(x;\bm{\lambda}+\bm{\delta}\bigr)
  =b_{n-1}(\bm{\lambda})\phi_n(x;\bm{\lambda}).
  \label{Adphi=bphi}
\end{align}
Here the coefficients $f_n(\bm{\lambda})$ and $b_{n-1}(\bm{\lambda})$
are the factors of
$\mathcal{E}_n(\bm{\lambda})=f_n(\bm{\lambda})b_{n-1}(\bm{\lambda})$:
\begin{equation}
  f_n(\bm{\lambda})\eqdef\left\{
  \begin{array}{ll}
  -2(n+g+h)&:\text{J}\\
  -2&:\text{L}
  \end{array}\right.\!\!,
  \quad
  b_{n-1}(\bm{\lambda})\eqdef -2n\ :\text{J\,\&\,L}.
\end{equation}
The forward and backward shift operators, $\mathcal{F}(\bm{\lambda})$
and $\mathcal{B}(\bm{\lambda})$, are defined in the following way and
they can be expressed in terms of $\eta$ only \cite{hos}:
\begin{align}
  \mathcal{F}(\bm{\lambda})&\eqdef
  \phi_0(x;\bm{\lambda}+\bm{\delta})^{-1}\circ
  \mathcal{A}(\bm{\lambda})\circ\phi_0(x;\bm{\lambda})
  =\frac{\phi_0(x;\bm{\lambda})}
  {\phi_0(x;\bm{\lambda}+\bm{\delta})}\,\frac{d}{dx}
  =\cF\frac{d}{d\eta},
  \label{Fdef}\\
  \mathcal{B}(\bm{\lambda})&\eqdef
  \phi_0(x;\bm{\lambda})^{-1}\circ
  \mathcal{A}(\bm{\lambda})^{\dagger}
  \circ\phi_0(x;\bm{\lambda}+\bm{\delta})\n
  &=-\frac{\phi_0(x;\bm{\lambda}+\bm{\delta})}{\phi_0(x;\bm{\lambda})}
  \Bigl(\frac{d}{dx}+\partial_x\bigl(w_0(x;\bm{\lambda})
  +w_0(x;\bm{\lambda}+\bm{\delta})\bigr)\Bigr)\n
  &=-4\cF^{-1}c_2(\eta)\Bigl(\frac{d}{d\eta}
  +\frac{c_1(\eta,\bm{\lambda})}{c_2(\eta)}\Bigr),
  \label{Bdef}
\end{align}
where $\cF$, $c_1(\eta,\bm{\lambda})$ and $c_2(\eta)$ are
\begin{equation}
  \cF\eqdef\left\{
  \begin{array}{ll}
  -4&\!:\text{J}\\
  2&\!:\text{L}
  \end{array}\right.\!\!,
  \ \ c_1(\eta,\bm{\lambda})\eqdef\left\{
  \begin{array}{ll}
  h-g-(g+h+1)\eta&\!:\text{J}\\
  g+\tfrac12-\eta&\!:\text{L}
  \end{array}\right.\!\!,
  \ \ c_2(\eta)\eqdef\left\{
  \begin{array}{ll}
  1-\eta^2&\!:\text{J}\\
  \eta&\!:\text{L}
  \end{array}\right.\!\!.\!\!
  \label{cF,c1,c2}
\end{equation}
Their action on the polynomial is
\begin{align}
  &\mathcal{F}(\bm{\lambda})P_n(\eta;\bm{\lambda})
  =f_n(\bm{\lambda})P_{n-1}(\eta;\bm{\lambda}+\bm{\delta}),
  \label{FP=fP}\\
  &\mathcal{B}(\bm{\lambda})P_{n-1}(\eta;\bm{\lambda}+\bm{\delta})
  =b_{n-1}(\bm{\lambda})P_n(\eta;\bm{\lambda}).
  \label{BP=bP}
\end{align}
These forward and backward shift relations are the factors of the second
order differential equations for the polynomial $P_n$ :
\begin{align}
  &\cF\partial_{\eta}P_n(\eta;\bm{\lambda})
  =f_n(\bm{\lambda})P_{n-1}(\eta;\bm{\lambda}+\bm{\delta}),
  \label{Pnforward}\\
  &c_1(\eta,\bm{\lambda})P_{n-1}(\eta;\bm{\lambda}+\bm{\delta})
  +c_2(\eta)\partial_{\eta}P_{n-1}(\eta;\bm{\lambda}+\bm{\delta})
  =-\tfrac14\cF b_{n-1}(\bm{\lambda})P_n(\eta;\bm{\lambda}),
  \label{Pnbackward}\\
  &c_2(\eta)\partial_{\eta}^2P_n(\eta;\bm{\lambda})
  +c_1(\eta,\bm{\lambda})\partial_{\eta}P_n(\eta;\bm{\lambda})
  =-\tfrac14\mathcal{E}_n(\bm{\lambda})P_n(\eta;\bm{\lambda}),
  \label{Pndiffeq}
\end{align}
which correspond to the properties of the (confluent) hypergeometric
function ${}_2F_1$ \eqref{2F1prop1}--\eqref{2F1prop3} or ${}_1F_1$
\eqref{1F1prop1}--\eqref{1F1prop3}, respectively.

The orthogonality reads
\begin{align}
  &\int_{x_1}^{x_2}\!\!\phi_{0}(x;\bm{\lambda})^2\,
  P_n(\eta(x);\bm{\lambda})P_m(\eta(x);\bm{\lambda})dx
  =h_n(\bm{\lambda})\delta_{nm},
  \label{intPnPm}\\
  &\qquad\quad h_n(\bm{\lambda})\eqdef\left\{
  \begin{array}{ll}
  {\displaystyle\frac{\Gamma(n+g+\frac12)\Gamma(n+h+\frac12)}
  {2\,n!(2n+g+h)\Gamma(n+g+h)}}&:\text{J}\\[10pt]
  \frac{1}{2\,n!}\Gamma(n+g+\frac12)&:\text{L}
  \end{array}\right.\!\!.
  \label{hn}
\end{align}

\section{Deforming function with continuous $\ell$}
\setcounter{equation}{0}

In deriving the Hamiltonians of the exceptional Jacobi and Laguerre
polynomials \cite{os16,os19}, the original system is deformed in terms
of a degree $\ell=1,2,\ldots$ polynomial $\xi_{\ell}(\eta;\bm{\lambda})$,
which is the eigenpolynomial (Jacobi or Laguerre) with twisted parameters.
We consider a real positive number $\ell$ instead of an integer.

Let us define the following deforming function
$\xi_{\ell}(\eta;\bm{\lambda})$ with $\ell\in\mathbb{R}_{>0}$:
\begin{align}
  \text{J1}:\ \xi_{\ell}(\eta;\bm{\lambda})&\eqdef
  \frac{\Gamma(g+2\ell-\frac12)}{\Gamma(\ell+1)\Gamma(g+\ell-\frac12)}\,
  {}_2F_1\Bigl(\genfrac{}{}{0pt}{}{-\ell,g-h+\ell-1}{g+\ell-\frac12}
  \Bigm|\frac{1-\eta}{2}\Bigr)
  \label{xilJ1}\\
  &=\frac{\Gamma(g+2\ell-\frac12)}{\Gamma(\ell+1)\Gamma(g+\ell-\frac12)}\,
  \Bigl(\frac{1+\eta}{2}\Bigr)^{h+\ell+\frac12}
  {}_2F_1\Bigl(\genfrac{}{}{0pt}{}{g+2\ell-\frac12,h+\frac12}{g+\ell-\frac12}
  \Bigm|\frac{1-\eta}{2}\Bigr),
  \label{xilJ1_2}\\
  \text{L1}:\ \xi_{\ell}(\eta;\bm{\lambda})&\eqdef
  \frac{\Gamma(g+2\ell-\frac12)}{\Gamma(\ell+1)\Gamma(g+\ell-\frac12)}\,
  {}_1F_1\Bigl(\genfrac{}{}{0pt}{}{-\ell}{g+\ell-\frac12}\Bigm|-\eta\Bigr)
  \label{xilL1}\\
  &=\frac{\Gamma(g+2\ell-\frac12)}{\Gamma(\ell+1)\Gamma(g+\ell-\frac12)}\,
  e^{-\eta}{}_1F_1\Bigl(\genfrac{}{}{0pt}{}{g+2\ell-\frac12}
  {g+\ell-\frac12}\Bigm|\eta\Bigr),
  \label{xilL1_2}
\end{align}
where the Kummer's transformation formula is used in the second equalities.
In addition to the condition $g,h>0$ \eqref{eta,lambda},
we restrict the parameters as follows:
\begin{equation}
  \left\{
  \begin{array}{ll}
  g>\tfrac32,\ h>\tfrac12&:\text{J1}\\[4pt]
  g>\tfrac32,&:\text{L1}
  \end{array}\right.\!\!,
  \label{pararange}
\end{equation}
then the deforming function $\xi_{\ell}(\eta(x);\bm{\lambda})$ has no zero
in the domain $x_1< x<x_2$.
This can be easily verified by using the power series definition of the
(confluent) hypergeometric function \eqref{2F1def},\eqref{1F1def} and
the alternative expressions \eqref{xilJ1_2} and \eqref{xilL1_2}.

Since the Jacobi and Laguerre polynomials are expressed as
\begin{align}
  P_n^{(\alpha,\beta)}(x)&=\frac{(\alpha+1)_n}{n!}\,
  {}_2F_1\Bigl(\genfrac{}{}{0pt}{}{-n,n+\alpha+\beta+1}{\alpha+1}\Bigm|
  \frac{1-x}{2}\Bigr),
  \label{Jac=2F1}\\
  L_n^{(\alpha)}(x)&=\frac{(\alpha+1)_n}{n!}\,
  {}_1F_1\Bigl(\genfrac{}{}{0pt}{}{-n}{\alpha+1}\Bigm|x\Bigr),
  \label{Lag=1F1}
\end{align}
this deforming function reduces to the deforming polynomial in
\cite{os16,os19} for integer $\ell$
\begin{equation}
  \ell\in\mathbb{Z}_{> 0}\ \ \Rightarrow
  \ \ \xi_{\ell}(\eta;\bm{\lambda})=\left\{
  \begin{array}{ll}
  P_{\ell}^{(g+\ell-\frac32,-h-\ell-\frac12)}(\eta)&:\text{J1}\\
  L_{\ell}^{(g+\ell-\frac32)}(-\eta)&:\text{L1}
  \end{array}\right.\!\!.
\end{equation}
We remark that we had restricted $g>h>0$ for J1 ($h>g>0$ for J2) in
\cite{os16,os19,hos} for a positive integer $\ell$, but this restriction
is unnecessary due to \eqref{xilJ1_2} and \eqref{xilL1_2}.

\bigskip

Here we present three formulas satisfied by the deforming function
$\xi_{\ell}(\eta;\bm{\lambda})$ \eqref{xildiffeq}--\eqref{xil(l)},
which will play important roles in the derivation of various results
in \S\,\ref{sec:deformed_system}:
\begin{align}
  &c_2(\eta)\partial_{\eta}^2\xi_{\ell}(\eta;\bm{\lambda})
  +\tilde{c}_1(\eta,\bm{\lambda},\ell)
  \partial_{\eta}\xi_{\ell}(\eta;\bm{\lambda})
  =-\tfrac14\widetilde{\mathcal{E}}_{\ell}(\bm{\lambda})
  \xi_{\ell}(\eta;\bm{\lambda}),
  \label{xildiffeq}\\
  &d_1(\bm{\lambda}+\ell\bm{\delta})\xi_{\ell}(\eta;\bm{\lambda})
  +d_2(\eta)\partial_{\eta}\xi_{\ell}(\eta;\bm{\lambda})
  =d_1(\bm{\lambda})\xi_{\ell}(\eta;\bm{\lambda}+\bm{\delta}),
  \label{xil(l+d)}\\
  &d_3(\bm{\lambda},\ell)\xi_{\ell}(\eta;\bm{\lambda}+\bm{\delta})
  +\frac{c_2(\eta)}{d_2(\eta)}\,
  \partial_{\eta}\xi_{\ell}(\eta;\bm{\lambda}+\bm{\delta})
  =d_3(\bm{\lambda}+\ell\bm{\delta},\ell)\xi_{\ell}(\eta;\bm{\lambda}),
  \label{xil(l)}
\end{align}
where $\tilde{c}_1(\eta,\bm{\lambda},\ell)$, $d_1(\bm{\lambda})$,
$d_2(\eta)$, $d_3(\bm{\lambda},\ell)$ and
$\widetilde{\mathcal{E}}_{\ell}(\bm{\lambda})$ are given by \cite{hos}
\begin{align}
  &\tilde{c}_1(\eta,\bm{\lambda},\ell)\eqdef\left\{
  \begin{array}{ll}
  -\bigl(g+h+2\ell-1+(g-h)\eta\bigr)&:\text{J1}\\
  g+\ell-\tfrac12+\eta&:\text{L1}
  \end{array}\right.\!\!,
  \label{c1t}\\
  &d_1(\bm{\lambda})\eqdef\left\{
  \begin{array}{ll}
  h+\frac12&:\text{J1}\\[2pt]
  1&:\text{L1}\\
  \end{array}\right.\!\!,
  \quad
  d_2(\eta)\eqdef\left\{
  \begin{array}{ll}
 -(1+\eta)&:\text{J1}\\
  1&:\text{L1}
  \end{array}\right.\!\!,
  \label{d1,d2}\\
  &d_3(\bm{\lambda},\ell)\eqdef
  g+\ell-\tfrac12\ :\text{J1\,\&\,L1},
  \quad
  \widetilde{\mathcal{E}}_{\ell}(\bm{\lambda})\eqdef\left\{
  \begin{array}{ll}
  4\ell(\ell+g-h-1)&:\text{J1}\\
  -4\ell&:\text{L1}
  \end{array}\right.\!\!.
\end{align}
The first equation \eqref{xildiffeq} is the differential equation for
the deforming function.
The eqs.\,\eqref{xil(l+d)}--\eqref{xil(l)} are identities relating
$\xi_{\ell}(\eta;\bm{\lambda})$ and
$\xi_{\ell}(\eta;\bm{\lambda}+\bm{\delta})$.
The eqs.\,\eqref{xildiffeq}--\eqref{xil(l)} are obtained from
the properties of the hypergeometric function
\eqref{2F1prop1}--\eqref{2F1prop5} and
\eqref{1F1prop1}--\eqref{1F1prop5}.
It is interesting to note that \eqref{xildiffeq} can be considered as
a consequence of \eqref{xil(l+d)} and \eqref{xil(l)}.

\bigskip

In the rest of this section we present some properties of the
hypergeometric functions ${}_2F_1$ and the confluent one ${}_1F_1$.
We assume that parameters ($a,b,c$) are generic.

The hypergeometric function ${}_2F_1$ is defined by
\begin{equation}
  {}_2F_1\Bigl(\genfrac{}{}{0pt}{}{a,b}{c}\Bigm|x\Bigr)
  \eqdef\sum_{k=0}^{\infty}
  \frac{(a)_k(b)_k}{(c)_k}\frac{x^k}{k!}\,,\quad(|x|<1).
  \label{2F1def}
\end{equation}
The following properties can be verified elementarily based on
\eqref{2F1def}:
\begin{align}
  &\frac{d}{dx}\,{}_2F_1\Bigl(\genfrac{}{}{0pt}{}{a,b}{c}\Bigm|x\Bigr)
  =\frac{ab}{c}\,
  {}_2F_1\Bigl(\genfrac{}{}{0pt}{}{a+1,b+1}{c+1}\Bigm|x\Bigr),
  \label{2F1prop1}\\
  &\Bigl(x(1-x)\frac{d}{dx}+c-(a+b+1)x\Bigr)
  {}_2F_1\Bigl(\genfrac{}{}{0pt}{}{a+1,b+1}{c+1}\Bigm|x\Bigr)
  =c\,{}_2F_1\Bigl(\genfrac{}{}{0pt}{}{a,b}{c}\Bigm|x\Bigr),
  \label{2F1prop2}\\
  &\Bigl(x(1-x)\frac{d^2}{dx^2}+\bigl(c-(a+b+1)x\bigr)\frac{d}{dx}-ab\Bigr)
  {}_2F_1\Bigl(\genfrac{}{}{0pt}{}{a,b}{c}\Bigm|x\Bigr)=0,
  \label{2F1prop3}\\
  &(a+b-c)\,{}_2F_1\Bigl(\genfrac{}{}{0pt}{}{a,b}{c}\Bigm|x\Bigr)
  +\frac{(c-a)(c-b)}{c}\,
  {}_2F_1\Bigl(\genfrac{}{}{0pt}{}{a,b}{c+1}\Bigm|x\Bigr)
  =(1-x)\frac{ab}{c}\,
  {}_2F_1\Bigl(\genfrac{}{}{0pt}{}{a+1,b+1}{c+1}\Bigm|x\Bigr),
  \label{2F1prop4}\\
  &{}_2F_1\Bigl(\genfrac{}{}{0pt}{}{a,b}{c}\Bigm|x\Bigr)
  -{}_2F_1\Bigl(\genfrac{}{}{0pt}{}{a,b}{c+1}\Bigm|x\Bigr)
  =\frac{x}{c}\frac{ab}{c+1}\,
  {}_2F_1\Bigl(\genfrac{}{}{0pt}{}{a+1,b+1}{c+2}\Bigm|x\Bigr).
  \label{2F1prop5}
\end{align}

The confluent hypergeometric function ${}_1F_1$ is defined by
\begin{equation}
  {}_1F_1\Bigl(\genfrac{}{}{0pt}{}{a}{b}\Bigm|x\Bigr)
  \eqdef\sum_{k=0}^{\infty}
  \frac{(a)_k}{(b)_k}\frac{x^k}{k!}\,.
  \label{1F1def}
\end{equation}
The following properties can be verified elementarily based on
\eqref{1F1def}:
\begin{align}
  &\frac{d}{dx}\,{}_1F_1\Bigl(\genfrac{}{}{0pt}{}{a}{b}\Bigm|x\Bigr)
  =\frac{a}{b}\,
  {}_1F_1\Bigl(\genfrac{}{}{0pt}{}{a+1}{b+1}\Bigm|x\Bigr),
  \label{1F1prop1}\\
  &\Bigl(x\frac{d}{dx}+b-x\Bigr)
  {}_1F_1\Bigl(\genfrac{}{}{0pt}{}{a+1}{b+1}\Bigm|x\Bigr)
  =b\,{}_1F_1\Bigl(\genfrac{}{}{0pt}{}{a}{b}\Bigm|x\Bigr),
  \label{1F1prop2}\\
  &\Bigl(x\frac{d^2}{dx^2}+(b-x)\frac{d}{dx}-a\Bigr)
  {}_1F_1\Bigl(\genfrac{}{}{0pt}{}{a}{b}\Bigm|x\Bigr)=0,
  \label{1F1prop3}\\
  &{}_1F_1\Bigl(\genfrac{}{}{0pt}{}{a}{b}\Bigm|x\Bigr)
  +\frac{a-b}{b}\,{}_1F_1\Bigl(\genfrac{}{}{0pt}{}{a}{b+1}\Bigm|x\Bigr)
  =\frac{a}{b}\,{}_1F_1\Bigl(\genfrac{}{}{0pt}{}{a+1}{b+1}\Bigm|x\Bigr),
  \label{1F1prop4}\\
  &{}_1F_1\Bigl(\genfrac{}{}{0pt}{}{a+1}{b}\Bigm|x\Bigr)
  -{}_1F_1\Bigl(\genfrac{}{}{0pt}{}{a}{b}\Bigm|x\Bigr)
  =\frac{x}{b}\,{}_1F_1\Bigl(\genfrac{}{}{0pt}{}{a+1}{b+1}\Bigm|x\Bigr).
  \label{1F1prop5}
\end{align}

\bigskip
To sum up, the deforming function $\xi_{\ell}(\eta(x);\bm{\lambda})$
possesses two important properties;
(\romannumeral1) it has no zero in the domain $x_1< x<x_2$,
(\romannumeral2) it satisfies the three formulas
\eqref{xildiffeq}--\eqref{xil(l)},
which are the essential properties of the deforming polynomials in the
theory of the exceptional Jacobi and Laguerre polynomials \cite{os19,hos}.

Before closing this section, let us briefly comment on the possible
``continuous $\ell$ versions" corresponding to the J2 Jacobi and L2
Laguerre polynomials.
The obvious candidates for the deforming function are:
\begin{align}
  \text{J2}:\ \xi_{\ell}(\eta;\bm{\lambda})&\eqdef
  \frac{\Gamma(-g+\frac12)}{\Gamma(\ell+1)\Gamma(-g-\ell+\frac12)}\,
  {}_2F_1\Bigl(\genfrac{}{}{0pt}{}{-\ell,h-g+\ell-1}{-g-\ell+\frac12}
  \Bigm|\frac{1-\eta}{2}\Bigr),
  \label{xilJ2}\\
  \text{L2}:\ \xi_{\ell}(\eta;\bm{\lambda})&\eqdef
  \frac{\Gamma(-g+\frac12)}{\Gamma(\ell+1)\Gamma(-g-\ell+\frac12)}\,
  {}_1F_1\Bigl(\genfrac{}{}{0pt}{}{-\ell}{-g-\ell+\frac12}\Bigm|\eta\Bigr).
  \label{xilL2}
\end{align}
For the above choices and a few other related candidates, we have not
been able to find proper parameter ranges in which the above two
properties (\romannumeral1) and (\romannumeral2) are satisfied and
at the same time invariant under the shifts, $g\to g+1$, $h\to h+1$,
so that the shape invariance method is applicable.
In other words, the J2 and L2 deformations are valid only for integer $\ell$.

\section{Deformed systems and intertwining relations}
\label{sec:deformed_system}
\setcounter{equation}{0}

We deform the original systems in terms of the deforming function
$\xi_{\ell}(\eta(x);\bm{\lambda})$ in exactly the same manner as in the
theory of the exceptional orthogonal Jacobi and Laguerre polynomials
\cite{os16,os19,hos}.

\subsection{Deformed systems}
\label{sec:deformed.sys}

For a real positive number $\ell$, we define a deformed Hamiltonian:
\begin{align}
  &\mathcal{H}_{\ell}(\bm{\lambda})\eqdef
  \mathcal{A}_{\ell}(\bm{\lambda})^{\dagger}
  \mathcal{A}_{\ell}(\bm{\lambda}),
  \label{deformham}\\
  &\mathcal{A}_{\ell}(x;\bm{\lambda})\eqdef
  \frac{d}{dx}-\partial_xw_{\ell}(x;\bm{\lambda}),\quad
  \mathcal{A}_{\ell}(x;\bm{\lambda})^{\dagger}=
  -\frac{d}{dx}-\partial_xw_{\ell}(x;\bm{\lambda}),\\
  &w_{\ell}(x;\bm{\lambda})\eqdef
  w_0(x;\bm{\lambda}+\ell\bm{\delta})
  +\log\frac{\xi_{\ell}(\eta(x);\bm{\lambda}+\bm{\delta})}
  {\xi_{\ell}(\eta(x);\bm{\lambda})}.
  \label{wl}
\end{align}
The overall normalisation of the deforming function $\xi_{\ell}$ is
immaterial for the deformation and the original Hamiltonian corresponds
to $\ell=0$.
This system is shape invariant,
\begin{equation}
  \mathcal{A}_{\ell}(\bm{\lambda})\mathcal{A}_{\ell}(\bm{\lambda})^{\dagger}
  =\mathcal{A}_{\ell}(\bm{\lambda+\bm{\delta}})^{\dagger}
  \mathcal{A}_{\ell}(\bm{\lambda}+\bm{\delta})
  +\mathcal{E}_{\ell,1}(\bm{\lambda}),
  \label{Xlshapeinv}
\end{equation}
or equivalently,
\begin{equation}
  \bigl(\partial_xw_{\ell}(x;\bm{\lambda})\bigr)^2
  -\partial_x^2w_{\ell}(x;\bm{\lambda})
  =\bigl(\partial_xw_{\ell}(x;\bm{\lambda}+\bm{\delta})\bigr)^2
  +\partial_x^2w_{\ell}(x;\bm{\lambda}+\bm{\delta})
  +\mathcal{E}_{\ell,1}(\bm{\lambda}).
  \label{Xlshapeinv2}
\end{equation}
As in the theory of the exceptional orthogonal Jacobi and Laguerre
polynomials \cite{os18}, this relation is reduced to an identity
involving cubic products of $\xi_{\ell}$, which can be proved by using
the three formulas \eqref{xildiffeq}--\eqref{xil(l)}.
The shape invariance determines the whole spectrum and eigenfunctions in
terms of the first excited state energy and the ground state wavefunction.
Eqs.\,\eqref{Alphiln=flnphiln}--\eqref{fln,bln} and
\eqref{FlPln=flnPln}--\eqref{BlPln=blnPln} are the consequences of
the shape invariance and the normalization of the eigenfunctions.

We present various properties of these deformed systems, which will be
derived without using shape invariance in the next subsection.
The Schr\"{o}dinger equation of this deformed system is
\begin{align}
  &\mathcal{H}_{\ell}(\bm{\lambda})\phi_{\ell,n}(x;\bm{\lambda})
  =\mathcal{E}_{\ell,n}(\bm{\lambda})\phi_{\ell,n}(x;\bm{\lambda})\quad
  (n=0,1,2,\ldots),\\
  &\phi_{\ell,n}(x;\bm{\lambda})
  =\psi_{\ell}(x;\bm{\lambda})P_{\ell,n}(\eta(x);\bm{\lambda}),\quad
  \psi_{\ell}(x;\bm{\lambda})\eqdef
  \frac{\phi_0(x;\bm{\lambda}+\ell\bm{\delta})}
  {\xi_{\ell}(\eta(x);\bm{\lambda})}.
\end{align}
The spectrum and the main part of the eigenfunction are
\begin{align}
  &\mathcal{E}_{\ell,n}(\bm{\lambda})
  =\mathcal{E}_n(\bm{\lambda}+\ell\bm{\delta}),\\
  &P_{\ell,n}(\eta;\bm{\lambda})\eqdef
  \frac{2}{\hat{f}_{\ell,n}(\bm{\lambda})}
  \Bigl(d_2(\eta)\xi_{\ell}(\eta;\bm{\lambda})\partial_{\eta}
  P_n(\eta;\bm{\lambda}+\ell\bm{\delta}+\bm{\tilde{\delta}})\n
  &\phantom{P_{\ell,n}(\eta;\bm{\lambda})
  =\frac{2}{\hat{f}_{\ell,n}(\bm{\lambda})}}
  -d_1(\bm{\lambda})\xi_{\ell}(\eta;\bm{\lambda}+\bm{\delta})
  P_n(\eta;\bm{\lambda}+\ell\bm{\delta}+\bm{\tilde{\delta}})\Bigr),
  \label{Pln}
\end{align}
where $\hat{f}_{\ell,n}(\bm{\lambda})$ and $\bm{\tilde{\delta}}$ will be
given in \eqref{hatfhatb}--\eqref{deltat}.
The Sturm--Liouville's theorem ensures that the function
$P_{\ell,n}(\eta(x);\bm{\lambda})$ has $n$ zeros in the domain
$x_1<x<x_2$.
The action of $\mathcal{A}_{\ell}(\bm{\lambda})$ and
$\mathcal{A}_{\ell}(\bm{\lambda})^{\dagger}$ on the eigenfunction is
\begin{align}
  &\mathcal{A}_{\ell}(\bm{\lambda})\phi_{\ell,n}(x;\bm{\lambda})
  =f_{\ell,n}(\bm{\lambda})
  \phi_{\ell,n-1}\bigl(x;\bm{\lambda}+\bm{\delta}\bigr),
  \label{Alphiln=flnphiln}\\
  &\mathcal{A}_{\ell}(\bm{\lambda})^{\dagger}
  \phi_{\ell,n-1}\bigl(x;\bm{\lambda}+\bm{\delta}\bigr)
  =b_{\ell,n-1}(\bm{\lambda})\phi_{\ell,n}(x;\bm{\lambda}),
  \label{Aldphiln=blnphiln}\\
  &f_{\ell,n}(\bm{\lambda})=f_n(\bm{\lambda}+\ell\bm{\delta}),\quad
  b_{\ell,n-1}(\bm{\lambda})=b_{n-1}(\bm{\lambda}+\ell\bm{\delta}).
  \label{fln,bln}
\end{align}
The forward and backward shift operators are defined in a similar way
as before
\begin{align}
  \mathcal{F}_{\ell}(\bm{\lambda})&\eqdef
  \psi_{\ell}(x;\bm{\lambda}+\bm{\delta})^{-1}\circ
  \mathcal{A}_{\ell}(\bm{\lambda})\circ\psi_{\ell}(x;\bm{\lambda})\n
  &=\cF\frac{\xi_{\ell}(\eta;\bm{\lambda}+\bm{\delta})}
  {\xi_{\ell}(\eta;\bm{\lambda})}\Bigl(\frac{d}{d\eta}
  -\partial_{\eta}\log\xi_{\ell}(\eta;\bm{\lambda}+\bm{\delta})\Bigr),
  \label{Fldef}\\
  \mathcal{B}_{\ell}(\bm{\lambda})&\eqdef
  \psi_{\ell}(x;\bm{\lambda})^{-1}\circ
  \mathcal{A}_{\ell}(\bm{\lambda})^{\dagger}
  \circ\psi_{\ell}(x;\bm{\lambda}+\bm{\delta})\n
  &=-4\cF^{-1}c_2(\eta)\frac{\xi_{\ell}(\eta;\bm{\lambda})}
  {\xi_{\ell}(\eta;\bm{\lambda}+\bm{\delta})}
  \Bigl(\frac{d}{d\eta}
  +\frac{c_1(\eta,\bm{\lambda}+\ell\bm{\delta})}{c_2(\eta)}
  -\partial_{\eta}\log\xi_{\ell}(\eta;\bm{\lambda})\Bigr),
  \label{Bldef}
\end{align}
and their action on $P_{\ell,n}$ is
\begin{align}
  &\mathcal{F}_{\ell}(\bm{\lambda})P_{\ell,n}(\eta;\bm{\lambda})
  =f_{\ell,n}(\bm{\lambda})P_{\ell,n-1}(\eta;\bm{\lambda}+\bm{\delta}),
  \label{FlPln=flnPln}\\
  &\mathcal{B}_{\ell}(\bm{\lambda})P_{\ell,n-1}(\eta;\bm{\lambda}+\bm{\delta})
  =b_{\ell,n-1}(\bm{\lambda})P_{\ell,n}(\eta;\bm{\lambda}).
  \label{BlPln=blnPln}
\end{align}
The second order differential operator
$\widetilde{\mathcal{H}}_{\ell}(\bm{\lambda})$ acting on the functions
$P_{\ell,n}(\eta;\bm{\lambda})$ is defined by
\begin{align}
  &\widetilde{\mathcal{H}}_{\ell}(\bm{\lambda})\eqdef
  \mathcal{B}_{\ell}(\bm{\lambda})\mathcal{F}_{\ell}(\bm{\lambda})
  =\psi_{\ell}(x;\bm{\lambda})^{-1}\circ\mathcal{H}_{\ell}(\bm{\lambda})
  \circ\psi_{\ell}(x;\bm{\lambda})\n
  &\phantom{\widetilde{\mathcal{H}}_{\ell}(\bm{\lambda})}
  =-4\Bigl(c_2(\eta)\frac{d^2}{d\eta^2}
  +\bigl(c_1(\eta,\bm{\lambda}+\ell\bm{\delta})-2c_2(\eta)
  \partial_{\eta}\log\xi_{\ell}(\eta;\bm{\lambda})\bigr)
  \frac{d}{d\eta}\n
  &\phantom{\widetilde{\mathcal{H}}_{\ell}(\bm{\lambda})=-4\Bigl(}
  -2d_2(\eta)d_3(\bm{\lambda},\ell)
  \partial_{\eta}\log\xi_{\ell}(\eta;\bm{\lambda})
  -\tfrac14\,\widetilde{\mathcal{E}}_{\ell}(\bm{\lambda})\Bigr),
  \label{Hlt}\\
  &\widetilde{\mathcal{H}}_{\ell}(\bm{\lambda})
  P_{\ell,n}(\eta;\bm{\lambda})
  =\mathcal{E}_{\ell,n}(\bm{\lambda})
  P_{\ell,n}(\eta;\bm{\lambda}),
\end{align}
where we have used \eqref{xildiffeq}--\eqref{xil(l)} in \eqref{Hlt}.

The orthogonality reads
\begin{equation}
  \int_{x_1}^{x_2}\!\!\psi_{\ell}(x;\bm{\lambda})^2\,
  P_{\ell,n}(\eta(x);\bm{\lambda})P_{\ell,m}(\eta(x);\bm{\lambda})dx
  =h_{\ell,n}(\bm{\lambda})\delta_{nm}.
\end{equation}
The normalisation constant $h_{\ell,n}(\bm{\lambda})$ is related to
$h_n(\bm{\lambda})$ \eqref{hn} as
\begin{equation}
  h_{\ell,n}(\bm{\lambda})
  =\frac{\hat{b}_{\ell,n}(\bm{\lambda})}{\hat{f}_{\ell,n}(\bm{\lambda})}
  h_n(\bm{\lambda}+\ell\bm{\delta}+\bm{\tilde{\delta}})
  =\frac{\hat{b}_{\ell,n}(\bm{\lambda})}{\hat{f}_{\ell,n}(\bm{\lambda})}
  \frac{\hat{f}_{0,n}(\bm{\lambda}+\ell\bm{\delta})}
  {\hat{b}_{0,n}(\bm{\lambda}+\ell\bm{\delta})}
  h_n(\bm{\lambda}+\ell\bm{\delta}),
  \label{hln2}
\end{equation}
where $\hat{f}_{\ell,n}(\bm{\lambda})$, $\hat{b}_{\ell,n}(\bm{\lambda})$
and $\bm{\tilde{\delta}}$ are given by
\begin{align}
  &\hat{f}_{\ell,n}(\bm{\lambda})\eqdef-2\times\left\{
  \begin{array}{ll}
  n+h+\frac12&:\text{J1}\\
  1&:\text{L1}
  \end{array}\right.\!\!,
  \quad
  \hat{b}_{\ell,n}(\bm{\lambda})\eqdef
  -2(n+g+2\ell-\tfrac12)\ :\text{J1\,\&\,L1},
  \label{hatfhatb}\\
  &\qquad\qquad\qquad\qquad\qquad
  \bm{\tilde{\delta}}\eqdef\left\{
  \begin{array}{ll}
  (-1,1)&:\text{J1}\\
  -1&:\text{L1}
  \end{array}\right.\!\!.
  \label{deltat}
\end{align}
In the second equality of \eqref{hln2} we have used the explicit
expressions of $h_n(\bm{\lambda})$ \eqref{hn}.

\subsection{Intertwining relations}

\subsubsection{General setting}
\label{sec:general}

For well-defined operators $\hat{\mathcal{A}}_{\ell}(\bm{\lambda})$ and
$\hat{\mathcal{A}}_{\ell}(\bm{\lambda})^{\dagger}$, let us define
a pair of Hamiltonians $\hat{\mathcal{H}}_{\ell}^{(\pm)}(\bm{\lambda})$
\begin{equation}
  \hat{\mathcal{H}}_{\ell}^{(+)}(\bm{\lambda})\eqdef
  \hat{\mathcal{A}}_{\ell}(\bm{\lambda})^{\dagger}
  \hat{\mathcal{A}}_{\ell}(\bm{\lambda}),\quad
  \hat{\mathcal{H}}_{\ell}^{(-)}(\bm{\lambda})\eqdef
  \hat{\mathcal{A}}_{\ell}(\bm{\lambda})
  \hat{\mathcal{A}}_{\ell}(\bm{\lambda})^{\dagger},
  \label{H+-}
\end{equation}
and consider their Schr\"{o}dinger equations, that is, the eigenvalue
problems:
\begin{equation}
  \hat{\mathcal{H}}_{\ell}^{(\pm)}(\bm{\lambda})
  \hat{\phi}_{\ell,n}^{(\pm)}(x;\bm{\lambda})
  =\hat{\mathcal{E}}_{\ell,n}^{(\pm)}(\bm{\lambda})
  \hat{\phi}_{\ell,n}^{(\pm)}(x;\bm{\lambda})\quad
  (n=0,1,2,\ldots).
  \label{H+-Scheq}
\end{equation}
By definition, all the eigenfunctions must be square integrable.
Obviously the pair of Hamiltonians are intertwined:
\begin{align}
  &\hat{\mathcal{H}}_{\ell}^{(+)}(\bm{\lambda})
  \hat{\mathcal{A}}_{\ell}(\bm{\lambda})^{\dagger}
  =\hat{\mathcal{A}}_{\ell}(\bm{\lambda})^{\dagger}
  \hat{\mathcal{A}}_{\ell}(\bm{\lambda})
  \hat{\mathcal{A}}_{\ell}(\bm{\lambda})^{\dagger}
  =\hat{\mathcal{A}}_{\ell}(\bm{\lambda})^{\dagger}
  \hat{\mathcal{H}}_{\ell}^{(-)}(\bm{\lambda}),\\
  &\hat{\mathcal{A}}_{\ell}(\bm{\lambda})
  \hat{\mathcal{H}}_{\ell}^{(+)}(\bm{\lambda})
  =\hat{\mathcal{A}}_{\ell}(\bm{\lambda})
  \hat{\mathcal{A}}_{\ell}(\bm{\lambda})^{\dagger}
  \hat{\mathcal{A}}_{\ell}(\bm{\lambda})
  =\hat{\mathcal{H}}_{\ell}^{(-)}(\bm{\lambda})
  \hat{\mathcal{A}}_{\ell}(\bm{\lambda}).
\end{align}
If $\hat{\mathcal{A}}_{\ell}(\bm{\lambda})
\hat{\phi}_{\ell,n}^{(+)}(x;\bm{\lambda})\neq 0$ and
$\hat{\mathcal{A}}_{\ell}(\bm{\lambda})^{\dagger}
\hat{\phi}_{\ell,n}^{(-)}(x;\bm{\lambda})\neq 0$, then the two systems
are exactly iso-spectral and there is one-to-one correspondence between
the eigenfunctions:
\begin{align}
  &\hat{\mathcal{E}}_{\ell,n}^{(+)}(\bm{\lambda})
  =\hat{\mathcal{E}}_{\ell,n}^{(-)}(\bm{\lambda}),\\
  &\hat{\phi}_{\ell,n}^{(-)}(x;\bm{\lambda})\propto
  \hat{\mathcal{A}}_{\ell}(\bm{\lambda})
  \hat{\phi}_{\ell,n}^{(+)}(x;\bm{\lambda}),\quad
  \hat{\phi}_{\ell,n}^{(+)}(x;\bm{\lambda})\propto
  \hat{\mathcal{A}}_{\ell}(\bm{\lambda})^{\dagger}
  \hat{\phi}_{\ell,n}^{(-)}(x;\bm{\lambda}).
\end{align}
This situation is called `broken susy' case in the parlance of
supersymmetric quantum mechanics \cite{susyqm, junkroy}.
It should be stressed that in the ordinary setting of Crum's theorem,
the zero mode of $ \hat{\mathcal{A}}_{\ell}(\bm{\lambda})$ is the
groundstate of $\hat{\mathcal{H}}_{\ell}^{(+)}(\bm{\lambda})$.
In that case, $\hat{\mathcal{H}}_{\ell}^{(+)}(\bm{\lambda})$ and
$\hat{\mathcal{H}}_{\ell}^{(-)}(\bm{\lambda})$ are iso-spectral except
for the groundstate of $\hat{\mathcal{H}}_{\ell}^{(+)}(\bm{\lambda})$.

In the following we will present the explicit forms of the operators
$\hat{\mathcal{A}}_{\ell}(\bm{\lambda})$ and
$\hat{\mathcal{A}}_{\ell}(\bm{\lambda})^{\dagger}$,
which intertwine the original systems in \S\,\ref{sec:org.sys} and
the deformed systems in \S\,\ref{sec:deformed.sys}.

\subsubsection{Intertwining the original and the deformed systems}
\label{sec:intrel}

Here we demonstrate that the Hamiltonian systems of the original
polynomials reviewed in \S\,\ref{sec:org.sys} and the deformation
summarised in \S\,\ref{sec:deformed.sys} are intertwined by the
Darboux-Crum transformation. 

The intertwining operators $\hat{\mathcal{A}}_{\ell}(\bm{\lambda})$
and $\hat{\mathcal{A}}_{\ell}(\bm{\lambda})^{\dagger}$ are given by
\begin{align}
  &\hat{\mathcal{A}}_{\ell}(x;\bm{\lambda})\eqdef
  \frac{d}{dx}-\partial_x\hat{w}_{\ell}(x;\bm{\lambda}),\quad
  \hat{\mathcal{A}}_{\ell}(x;\bm{\lambda})^{\dagger}=
  -\frac{d}{dx}-\partial_x\hat{w}_{\ell}(x;\bm{\lambda}),\\
  &\hat{w}_{\ell}(x;\bm{\lambda})\eqdef
  \tilde{w}_0(x;\bm{\lambda}+\ell\bm{\delta})
  +\log\xi_{\ell}(\eta(x);\bm{\lambda}),
  \label{hatwl}\\
  &\tilde{w}_0(x;\bm{\lambda})\eqdef\left\{
  \begin{array}{ll}
  (g-1)\log\sin x-h\log\cos x&:\text{J1}\\[4pt]
  \frac{x^2}{2}+(g-1)\log x&:\text{L1}
  \end{array}\right.\!\!.
\end{align}
These have exactly the same form as those used for the exceptional J1
Jacobi and L1 Laguerre polynomials \cite{stz}. See also a similar work
\cite{gomez2}.
It is illuminating to compare these prepotential \eqref{hatwl} with
those of the original \eqref{w0} and deformed \eqref{wl} systems.
Again it is obvious that the overall normalisation of the deforming
polynomial $\xi_{\ell}$ is immaterial for
$\hat{\mathcal{A}}_{\ell}(\bm{\lambda})$ and
$\hat{\mathcal{A}}_{\ell}(\bm{\lambda})^{\dagger}$.

For this choice of $\hat{\mathcal{A}}_{\ell}(\bm{\lambda})$ and
$\hat{\mathcal{A}}_{\ell}(\bm{\lambda})^{\dagger}$, one of the pair of
Hamiltonians $\hat{\mathcal{H}}_{\ell}^{(+)}(\bm{\lambda})$
\eqref{H+-} becomes proportional to the original Hamiltonian
$\mathcal{H}(\bm{\lambda})$ \eqref{origham}
with $\bm{\lambda}\to\bm{\lambda}+\ell\bm{\delta}+\bm{\tilde{\delta}}$
and the partner Hamiltonian $\hat{\mathcal{H}}_{\ell}^{(-)}(\bm{\lambda})$
is proportional to the deformed Hamiltonian
$\mathcal{H}_{\ell}(\bm{\lambda})$ \eqref{deformham}, up to a common
additive constant:
\begin{align}
  \hat{\mathcal{H}}_{\ell}^{(+)}(\bm{\lambda})
  &=\mathcal{H}(\bm{\lambda}+\ell\bm{\delta}+\bm{\tilde{\delta}})
  +\hat{f}_{\ell,0}(\bm{\lambda})\hat{b}_{\ell,0}(\bm{\lambda}),
  \label{Hl+=H}\\
  \hat{\mathcal{H}}_{\ell}^{(-)}(\bm{\lambda})
  &=\mathcal{H}_{\ell}(\bm{\lambda})
  +\hat{f}_{\ell,0}(\bm{\lambda})\hat{b}_{\ell,0}(\bm{\lambda}).
  \label{Hl-=Hl}
\end{align}
These fundamental results can be obtained by explicit calculation, in
which the three formulas \eqref{xildiffeq}--\eqref{xil(l)} are used.

It is instructive to verify that the zero modes of
$\hat{\mathcal{A}}_{\ell}(\bm{\lambda})$ and
$\hat{\mathcal{A}}_{\ell}(\bm{\lambda})^{\dagger}$ do not belong to
the Hilbert space of the eigenfunctions.
In fact, the zero mode $\hat{\mathcal{A}}_{\ell}(\bm{\lambda})$ is
\begin{equation}
  \hat{\mathcal{A}}_{\ell}(\bm{\lambda})\chi=0,\quad
  \chi=e^{\hat{w}_{\ell}(x;\bm{\lambda})}
  =e^{\tilde{w}_0(x;\bm{\lambda}+\ell\bm{\delta})}
  \xi_{\ell}(\eta(x);\bm{\lambda}),
\end{equation}
which is non-square integrable for the chosen parameter range
\eqref{pararange}.
The zero mode of $\hat{\mathcal{A}}_{\ell}(\bm{\lambda})^{\dagger}$ is
\begin{equation}
  \hat{\mathcal{A}}_{\ell}(\bm{\lambda})^{\dagger}\rho=0,\quad
  \rho=e^{-\hat{w}_{\ell}(x;\bm{\lambda})}
  =\frac{e^{-\tilde{w}_0(x;\bm{\lambda}+\ell\bm{\delta})}}
  {\xi_{\ell}(\eta(x);\bm{\lambda})}=\frac{1}{\chi},
\end{equation}
which is also non-square integrable for the chosen parameter range
\eqref{pararange}.
Thus the `broken susy' case is demonstrated \cite{susyqm,junkroy}.

Based on the results \eqref{Hl+=H}--\eqref{Hl-=Hl}, we have
\begin{gather}
  \hat{\phi}_{\ell,n}^{(+)}(x;\bm{\lambda})
  =\phi_n(x;\bm{\lambda}+\ell\bm{\delta}+\bm{\tilde{\delta}}),\quad
  \hat{\phi}_{\ell,n}^{(-)}(x;\bm{\lambda})
  =\phi_{\ell,n}(x;\bm{\lambda}),
  \label{phi+-=..}\\
  \hat{\mathcal{E}}_{\ell,n}^{(\pm)}(\bm{\lambda})
  =\mathcal{E}_n(\bm{\lambda}+\ell\bm{\delta}+\bm{\tilde{\delta}})
  +\hat{f}_{\ell,0}(\bm{\lambda})\hat{b}_{\ell,0}(\bm{\lambda})
  =\mathcal{E}_{\ell,n}(\bm{\lambda})
  +\hat{f}_{\ell,0}(\bm{\lambda})\hat{b}_{\ell,0}(\bm{\lambda}).
  \label{E+-=..}
\end{gather}
Then it is trivial to verify $\hat{\mathcal{A}}_{\ell}(\bm{\lambda})
\hat{\phi}_{\ell,n}^{(+)}(x;\bm{\lambda})\neq 0$ and
$\hat{\mathcal{A}}_{\ell}(\bm{\lambda})^{\dagger}
\hat{\phi}_{\ell,n}^{(-)}(x;\bm{\lambda})\neq 0$.
For, if one of the eigenfunction is annihilated by
$\hat{\mathcal{A}}_{\ell}(\bm{\lambda})$
($\hat{\mathcal{A}}_{\ell}(\bm{\lambda})^\dagger$), the left hand side
of \eqref{Hl+=H}(\eqref{Hl-=Hl}) vanishes, whereas the right hand side
is $\mathcal{E}_n(\bm{\lambda}+\ell\bm{\delta}+\bm{\tilde{\delta}})+
\hat{f}_{\ell,0}(\bm{\lambda})\hat{b}_{\ell,0}(\bm{\lambda})$
times the eigenfunction, which is obviously non-vanishing.
Note that $\mathcal{E}_n(\bm{\lambda}+\ell\bm{\delta}+\bm{\tilde{\delta}})
=\mathcal{E}_n(\bm{\lambda}+\ell\bm{\delta})$.

The correspondence of the pair of eigenfunctions
$\hat{\phi}_{\ell,n}^{(\pm)}(x)$ is expressed as
\begin{equation}
  \hat{\phi}_{\ell,n}^{(-)}(x;\bm{\lambda})
  =\frac{\hat{\mathcal{A}}_{\ell}(\bm{\lambda})
  \hat{\phi}_{\ell,n}^{(+)}(x;\bm{\lambda})}
  {\hat{f}_{\ell,n}(\bm{\lambda})},
  \quad
  \hat{\phi}_{\ell,n}^{(+)}(x;\bm{\lambda})
  =\frac{\hat{\mathcal{A}}_{\ell}(\bm{\lambda})^{\dagger}
  \hat{\phi}_{\ell,n}^{(-)}(x;\bm{\lambda})}
  {\hat{b}_{\ell,n}(\bm{\lambda})}.
  \label{phi+<->phi-}
\end{equation}
Let us introduce operators $\hat{\mathcal{F}}_{\ell}(\bm{\lambda})$ and
$\hat{\mathcal{B}}_{\ell}(\bm{\lambda})$ defined by
\begin{align}
  \hat{\mathcal{F}}_{\ell}(\bm{\lambda})&\eqdef
  \psi_{\ell}(x;\bm{\lambda})^{-1}\circ
  \hat{\mathcal{A}}_{\ell}(\bm{\lambda})\circ
  \phi_0(x;\bm{\lambda}+\ell\bm{\delta}+\bm{\tilde{\delta}}),
  \label{hatFldef}\\
  \hat{\mathcal{B}}_{\ell}(\bm{\lambda})&\eqdef
  \phi_0(x;\bm{\lambda}+\ell\bm{\delta}+\bm{\tilde{\delta}})^{-1}\circ
  \hat{\mathcal{A}}_{\ell}(\bm{\lambda})^{\dagger}\circ
  \psi_{\ell}(x;\bm{\lambda}),
\end{align}
which can be expressed in terms of $\eta$:
\begin{align}
  \hat{\mathcal{F}}_{\ell}(\bm{\lambda})&=
  2\Bigl(d_2(\eta)\xi_{\ell}(\eta;\bm{\lambda})\frac{d}{d\eta}
  -d_1(\bm{\lambda})\xi_{\ell}(\eta;\bm{\lambda}+\bm{\delta})\Bigr),
  \label{hatFlform}\\
  \hat{\mathcal{B}}_{\ell}(\bm{\lambda})&=
  \frac{-2}{\xi_{\ell}(\eta;\bm{\lambda})}\Bigl(
  \frac{c_2(\eta)}{d_2(\eta)}\frac{d}{d\eta}+d_3(\bm{\lambda},\ell)\Bigr).
   \label{hatBlform}
\end{align}
The operators $\hat{\mathcal{F}}_{\ell}(\bm{\lambda})$ and
$\hat{\mathcal{B}}_{\ell}(\bm{\lambda})$ act as the forward and backward
shift operators connecting the original orthogonal polynomials $P_n(\eta)$
and the orthogonal functions $P_{\ell,n}(\eta)$:
\begin{align}
  \hat{\mathcal{F}}_{\ell}(\bm{\lambda})
  P_n(\eta;\bm{\lambda}+\ell\bm{\delta}+\bm{\tilde{\delta}})
  =\hat{f}_{\ell,n}(\bm{\lambda})P_{\ell,n}(\eta;\bm{\lambda}),
  \label{FhatPn=Pln}\\
  \hat{\mathcal{B}}_{\ell}(\bm{\lambda})P_{\ell,n}(\eta;\bm{\lambda})
  =\hat{b}_{\ell,n}(\bm{\lambda})
  P_n(\eta;\bm{\lambda}+\ell\bm{\delta}+\bm{\tilde{\delta}}).
  \label{BhatPln=Pn}
\end{align}
The former relation \eqref{FhatPn=Pln} with the explicit form of
$\hat{\mathcal{F}}_{\ell}(\bm{\lambda})$ \eqref{hatFlform} provides the
explicit expression \eqref{Pln} of the main part of the eigenfunction.
Other simple consequences of these relations are
\begin{equation}
  \hat{\mathcal{E}}_{\ell,n}^{(\pm)}(\bm{\lambda})
  =\hat{f}_{\ell,n}(\bm{\lambda})\hat{b}_{\ell,n}(\bm{\lambda}),\quad
  \mathcal{E}_n(\bm{\lambda}+\ell\bm{\delta})
  =\hat{f}_{\ell,n}(\bm{\lambda})\hat{b}_{\ell,n}(\bm{\lambda})
  -\hat{f}_{\ell,0}(\bm{\lambda})\hat{b}_{\ell,0}(\bm{\lambda}).
  \label{Eln+-}
\end{equation}

The operator $\hat{\mathcal{A}}_{\ell}(\bm{\lambda})$
intertwines those of the original and deformed systems
$\mathcal{A}(\bm{\lambda})$ and $\mathcal{A}_{\ell}(\bm{\lambda})$:
\begin{align}
  &\hat{\mathcal{A}}_{\ell}(\bm{\lambda}+\bm{\delta})
  \mathcal{A}(\bm{\lambda}+\ell\bm{\delta}+\bm{\tilde{\delta}})
  =\mathcal{A}_{\ell}(\bm{\lambda})
  \hat{\mathcal{A}}_{\ell}(\bm{\lambda}),
  \label{AhA=AlAh}\\
  &\hat{\mathcal{A}}_{\ell}(\bm{\lambda})
  \mathcal{A}(\bm{\lambda}+\ell\bm{\delta}+\bm{\tilde{\delta}})^{\dagger}
  =\mathcal{A}_{\ell}(\bm{\lambda})^{\dagger}
  \hat{\mathcal{A}}_{\ell}(\bm{\lambda}+\bm{\delta}).
  \label{AhAd=AldAh}
\end{align}
These relations can be obtained by explicit calculation, in which
the three formulas \eqref{xildiffeq}--\eqref{xil(l)} are used.
In terms of the definitions of the forward shift operators
$\mathcal{F}(\bm{\lambda})$ \eqref{Fdef},
$\mathcal{F}_{\ell}(\bm{\lambda})$ \eqref{Fldef},
$\hat{\mathcal{F}}_{\ell}(\bm{\lambda})$ \eqref{hatFldef}, and
the backward shift operators $\mathcal{B}(\bm{\lambda})$ \eqref{Bdef},
$\mathcal{B}_{\ell}(\bm{\lambda})$ \eqref{Bldef},
the above relations are rewritten as:
\begin{align}
  &\hat{\mathcal{F}}_{\ell}(\bm{\lambda}+\bm{\delta})
  \mathcal{F}(\bm{\lambda}+\ell\bm{\delta}+\bm{\tilde{\delta}})
  =\mathcal{F}_{\ell}(\bm{\lambda})
  \hat{\mathcal{F}}_{\ell}(\bm{\lambda}),
  \label{FlhF=FlFlh}\\
  &\hat{\mathcal{F}}_{\ell}(\bm{\lambda})
  \mathcal{B}(\bm{\lambda}+\ell\bm{\delta}+\bm{\tilde{\delta}})
  =\mathcal{B}_{\ell}(\bm{\lambda})
  \hat{\mathcal{F}}_{\ell}(\bm{\lambda}+\bm{\delta}).
  \label{FlhB=BlFlh}
\end{align}

By applying $\hat{\mathcal{A}}_{\ell}(\bm{\lambda}+\bm{\delta})$ and
$\hat{\mathcal{A}}_{\ell}(\bm{\lambda})$ to \eqref{Aphi=fphi} and
\eqref{Adphi=bphi} with a replacement
$\bm{\lambda}\to\bm{\lambda}+\ell\bm{\delta}+\bm{\tilde{\delta}}$
respectively, together with the use of \eqref{AhA=AlAh},
\eqref{AhAd=AldAh} and \eqref{phi+<->phi-}, we obtain
\begin{align}
  \mathcal{A}_{\ell}(\bm{\lambda})\phi_{\ell,n}(x;\bm{\lambda})
  &=f_n(\bm{\lambda}+\ell\bm{\delta}+\bm{\tilde{\delta}})
  \frac{\hat{f}_{\ell,n-1}(\bm{\lambda}+\bm{\delta})}
  {\hat{f}_{\ell,n}(\bm{\lambda})}\,
  \phi_{\ell,n-1}(x;\bm{\lambda}+\bm{\delta})\n
  &=f_n(\bm{\lambda}+\ell\bm{\delta})
  \phi_{\ell,n-1}(x;\bm{\lambda}+\bm{\delta}),
  \label{Alphiln=fnphiln2}\\
  \mathcal{A}_{\ell}(\bm{\lambda})^{\dagger}
  \phi_{\ell,n-1}(x;\bm{\lambda}+\bm{\delta})
  &=b_{n-1}(\bm{\lambda}+\ell\bm{\delta}+\bm{\tilde{\delta}})
  \frac{\hat{f}_{\ell,n}(\bm{\lambda})}
  {\hat{f}_{\ell,n-1}(\bm{\lambda}+\bm{\delta})}\,
  \phi_{\ell,n}(x;\bm{\lambda})\n
  &=b_{n-1}(\bm{\lambda}+\ell\bm{\delta})
  \phi_{\ell,n}(x;\bm{\lambda}+\bm{\delta}).
  \label{Aldphiln=bnphiln2}
\end{align}
In the calculation use is made of the explicit forms of
$\hat{f}_{\ell,n}(\bm{\lambda})$, $f_n(\bm{\lambda})$ and
$b_n(\bm{\lambda})$ in the second equalities.
This provides a proof of \eqref{Alphiln=flnphiln}--\eqref{fln,bln}
without recourse to the shape invariance.
Likewise the above intertwining relations of the forward-backward shift
operators \eqref{FlhF=FlFlh}--\eqref{FlhB=BlFlh} give a proof of
\eqref{FlPln=flnPln}--\eqref{BlPln=blnPln}, respectively, again without
recourse to the shape invariance.

Eq.\,\eqref{hln2} is shown in the following way:
\begin{align}
  &\quad\hat{f}_{\ell,n}(\bm{\lambda})\hat{f}_{\ell,m}(\bm{\lambda})
  \int_{x_1}^{x_2}dx\,\phi_{\ell,n}(x;\bm{\lambda})
  \phi_{\ell,m}(x;\bm{\lambda})\n
  &\stackrel{\text{(\romannumeral1)}}{=}
  \int_{x_1}^{x_2}dx\,\hat{\mathcal{A}}_{\ell}(\bm{\lambda})
  \phi_n(x;\bm{\lambda}+\ell\bm{\delta}+\bm{\tilde{\delta}})\cdot
  \hat{\mathcal{A}}_{\ell}(\bm{\lambda})
  \phi_m(x;\bm{\lambda}+\ell\bm{\delta}+\bm{\tilde{\delta}})\n
  &\stackrel{\text{(\romannumeral2)}}{=}
  \int_{x_1}^{x_2}dx\,\hat{\mathcal{A}}_{\ell}(\bm{\lambda})^{\dagger}
  \hat{\mathcal{A}}_{\ell}(\bm{\lambda})
  \phi_n(x;\bm{\lambda}+\ell\bm{\delta}+\bm{\tilde{\delta}})\cdot
  \phi_m(x;\bm{\lambda}+\ell\bm{\delta}+\bm{\tilde{\delta}})\n
  &\stackrel{\text{(\romannumeral3)}}{=}
  \hat{\mathcal{E}}^{(+)}_{\ell,n}(\bm{\lambda})\int_{x_1}^{x_2}dx\,
  \phi_n(x;\bm{\lambda}+\ell\bm{\delta}+\bm{\tilde{\delta}})
  \phi_m(x;\bm{\lambda}+\ell\bm{\delta}+\bm{\tilde{\delta}})\n
  &\stackrel{\text{(\romannumeral4)}}{=}
  \hat{f}_{\ell,n}(\bm{\lambda})\hat{b}_{\ell,n}(\bm{\lambda})
  h_n(\bm{\lambda}+\ell\bm{\delta}+\bm{\tilde{\delta}})\delta_{nm}.
  \label{intformula}
\end{align}
Here we have used
\eqref{phi+<->phi-} and \eqref{phi+-=..} in (\romannumeral1),
an integration by parts in (\romannumeral2),
\eqref{H+-Scheq} and \eqref{phi+-=..} in (\romannumeral3),
\eqref{Eln+-} and \eqref{intPnPm} in (\romannumeral4).

\subsection{Limit: Jacobi $\to$ Laguerre}

It is well known \cite{szego} that the Laguerre polynomial
$L_n^{(\alpha)}(x)$ is obtained from the Jacobi polynomial
$P_n^{(\alpha,\beta)}(x)$ in a limit:
\begin{equation}
  \lim_{\beta\to\infty}P_n^{(\alpha,\,\pm\beta)}
  \bigl(1-2x\beta^{-1}\bigr)
  =L_n^{(\alpha)}(\pm x).
  \label{JtoL}
\end{equation}
It is also known that the radial oscillator potential can be obtained
from the trigonometric DPT potential in the limit of infinite coupling
$h\to\infty$ together with the rescaling of the coordinate:
\begin{equation}
  x=\frac{x^{\text{L}}}{\sqrt{h}},\quad
  0<x<\frac{\pi}{2}\Longleftrightarrow 0<x^{\text{L}}<\frac{\pi}{2}\sqrt{h}\,.
  \label{rescale}
\end{equation}
The two prepotentials \eqref{w0} are related \cite{os19}:
\begin{align}
  &\eta^{\text{J}}(x)=1-2\eta^{\text{L}}(x^{\text{L}})h^{-1}+O(h^{-2}),\\
  &\lim_{h\to\infty}\bigl(w_0^{\text{J}}(x;g,h)+\tfrac12g\log h\bigr)
  =w_0^{\text{L}}(x^{\text{L}};g).
\end{align}

Here we will show that the above two ``continuous $\ell$ versions",
the J1 Jacobi and L1 Laguerre, are connected by the same limit.
By using the series definitions of the (confluent) hypergeometric
functions \eqref{2F1def} and \eqref{1F1def}, one obtains
\begin{align}
  &\lim_{h\to\infty}
  {}_2F_1\Bigl(\genfrac{}{}{0pt}{}{-\ell,g-h+\ell-1}{g+\ell-\frac12}
  \Bigm|\frac{1-\eta^{\text J}(x)}{2}\Bigr)=
   {}_1F_1\Bigl(\genfrac{}{}{0pt}{}{-\ell}{g+\ell-\frac12}
  \Bigm|-\eta^{\text L}(x^{\text L})\Bigr),\\[2pt]
  &\lim_{h\to\infty}\xi_{\ell}^{\text J}(\eta^{\text J}(x);g,h)
  =\xi_{\ell}^{\text L}(\eta^{\text L}(x^{\text L});g).
\end{align}
At the same time we have
\begin{equation}
  \lim_{h\to\infty}\bigl(\tilde{w}_0^{\text{J}}(x;g,h)
  +\tfrac12(g-1)\log h\bigr)
  =\tilde{w}_0^{\text{L}}(x^{\text{L}};g).
\end{equation}
Thus the limiting procedure connects the original systems, the deformed
systems and the intertwining relations.

\section{Summary and Comments}
\setcounter{equation}{0}

A new family of shape invariantly deformed Darboux-P\"oschl-Teller
potentials \cite{dpt} is presented. It is a ``continuous $\ell$ version"
of the potentials corresponding to the exceptional ($X_{\ell}$) Jacobi
polynomials \cite{os16,os19,os18,hos,stz,os17,os20}.
The method of deformation, intertwining relations, etc are almost
parallel with those in the theory of the exceptional orthogonal
polynomials, that is, for integer $\ell$.
In the well known limit leading from the Jacobi polynomials (the
hypergeometric function) to the Laguerre polynomials (the confluent
hypergeometric function), the family of shape invariantly deformed
radial oscillator potentials with continuous $\ell$ is obtained.
The latter is known as an example of `conditionally exactly solvable
potentials' \cite{junkroy}.
It should be stressed that the ``continuous $\ell$ version" of the
exceptional orthogonal polynomials exists only for the first type,
the J1 $X_{\ell}$ Jacobi and L1 $X_{\ell}$ Laguerre polynomials.

The Hamiltonian of the DPT (radial oscillator) potential is known to
have infinitely many non-singular factorisations, up to an additive
constant, related with the exceptional orthogonal ($X_{\ell}$) polynomials,
$\ell=1,2, \ldots$, \cite{stz}.
Now we have demonstrated that the same Hamiltonian, up to an additive
constant, allow non-singular factorisations \eqref{Hl+=H} parametrised
by a continuous real number $\ell>0$.

\bigskip

In discrete quantum mechanics \cite{os4,os13,os14}, a similar deformation
based on a deforming polynomial $\xi_{\ell}$ with integer
$\ell=1,2,\ldots$, was studied \cite{os17,os20}, in which the
exceptional continuous Hahn, Wilson and Askey-Wilson polynomials were
obtained. The deforming polynomial $\xi_{\ell}(\eta(x);\bm{\lambda})$
satisfies three formulas (2.67)--(2.69) in \cite{os20}, which would
correspond to \eqref{xildiffeq}--\eqref{xil(l)} in the present paper.
One naturally wonders if a similar ``continuous $\ell$ version" could
be constructed or not. The answer is negative. Let us explain by taking
the simplest example of the continuous Hahn case. The theory has two
parameters, $a_1$ and $a_2$, with $a_1>0$ and $\text{Re}(a_2)>0$.
The continuous Hahn polynomial is defined by the {\em terminating\/}
hypergeometric function
\begin{equation}
  p_n(x;a_1,a_2,a_1,a_2^*)\eqdef
  i^n\frac{(2a_1)_n(a_1+a_2^*)_n}{n!}\,
  {}_3F_2\Bigl(\genfrac{}{}{0pt}{}{-n,\,n+2a_1+a_2+a_2^*-1,\,a_1+ix}
  {2a_1,\,a_1+a_2^*}\!\Bigm|\!1\Bigr).
  \label{defcH}
\end{equation}
If one replaces $n$ by a real positive number $\ell$, the hypergeometric
series ${}_3F_2$ becomes {\em nonterminating\/} and divergent, for
$\text{Re}(a_2)>1$. See Theorem 2.1.2 in \cite{askey}.
Even for a very limited parameter range $0<\text{Re}(a_2)<\frac12$,
the hypergeometric series ${}_3F_2$ will be divergent after one step
of shape-invariant transformation $a_2\to a_2+\frac12$.
Note that the ``continuous $\ell$ version" implies
$a_2\to a_2+\frac12(\ell-1)$, which does not improve the situation.
Thus we conclude that the ``continuous $\ell$ version" of the $X_{\ell}$
continuous Hahn polynomials does not exist.
In contrast, the (basic) hypergeometric series ${}_4F_3$ (${}_4\phi_3$)
appearing in the definition of the Wilson (Askey-Wilson) polynomial will
not diverge for generic $\ell$. See Theorem 2.1.2 of \cite{askey} for the
${}_4F_3$ case. The convergence of ${}_4\phi_3$ in the Askey-Wilson case
is due to the $q^k$ factor.
However, no difference equation governing the nonterminating ${}_4F_3$
(${}_4\phi_3$) series corresponding to (2.67) in \cite{os20} is known.
Thus the ``continuous $\ell$ version" of the deformation is not possible.

\section*{Acknowledgements}

R.\,S. is supported in part by Grant-in-Aid for Scientific Research
from the Ministry of Education, Culture, Sports, Science and Technology
(MEXT), No.19540179.


\end{document}